\documentclass[aps,prl,twocolumn,groupedaddress,showpacs]{revtex4}
\usepackage{graphicx}% Include figure files
\usepackage{dcolumn}% Align table columns on decimal point
\usepackage{bm}% bold math

\begin{document}

\title{Fermionic functionals without Grassmann numbers} 

\author{Hrvoje Nikoli\'c}
\affiliation{Theoretical Physics Division, Rudjer Bo\v{s}kovi\'{c} Institute, 
P.O.B. 180, HR-10002 Zagreb, Croatia.}
\email{hrvoje@thphys.irb.hr}

\date{\today}

\begin{abstract}
Since any fermionic operator $\psi$ can be written as 
$\psi=q+ip$, where $q$ and $p$ are hermitian operators, we use the 
eigenvalues of $q$ and $p$ to construct a functional 
formalism for calculating matrix elements that involve 
fermionic fields.   
The formalism is similar to that for bosonic fields and 
does not involve Grassmann numbers. This makes possible to perform
numerical fermionic lattice computations that are  
much faster than not only other algorithms for fermions, but 
also algorithms for bosons.
 
\end{abstract}

\pacs{03.65.Db, 11.15.Tk}

\maketitle

Since $\hat{\psi}^2=0$ for a fermionic field operator $\hat{\psi}$, 
this operator does not possess nonzero c-number eigenvalues. Therefore, 
the functional formalism for fermions involves anticommuting 
Grassmann numbers \cite{berez}. This makes the notion of a 
classical fermion field conceptually difficult. 
This is also problematic from a practical point 
of view, because the functional integrals 
over Grassmann numbers cannot be 
performed numerically on a computer, which makes lattice calculations 
with dynamical fermions difficult. For typical Hamiltonians, the 
integration over fermionic fields can be reduced to the computation 
of the determinants of huge matrices, which costs a lot of computer 
time. Therefore,
most of the existing lattice calculations are based on a quenched 
approximation, in which fermions are not treated dynamically
(see, e.g., \cite{davies} and references therein). Various methods to 
improve and speed up the treatment of fermions on a lattice have been 
proposed, such as those in Refs.~\cite{duan,lus,slav,cre,bak}. 
(Ref.~\cite{bak} contains a longer list of references for various 
methods.) However, the efficiency of all these existing methods cannot be 
compared with the efficiency of methods for functional integration 
over bosonic fields \cite{cre2}.   

In this Letter, we construct an entirely new method of 
functional calculus for fermionic fields. It is based on the fact 
that any linear operator $\hat{\psi}$ can be written in a 
unique way as a sum of a hermitian and an antihermitian 
operator. In other words, it can be written as
$\hat{\psi}=\hat{q}+i\hat{p}$, where $\hat{q}$ and 
$\hat{p}$ are hermitian operators. Since hermitian operators 
can always be diagonalized, we can use the eigenvalues and 
eigenstates of $\hat{q}$ and $\hat{p}$ to construct a functional 
formalism similar to that for bosons. We shall see that 
$\hat{q}$ and $\hat{p}$ play a role similar to that
of the canonical coordinate 
and the canonical momentum for bosons. This also suggests that 
$\hat{q}$ and $\hat{p}$ can be naturally interpreted as observables 
for fermionic fields, which may be relevant for a deeper understanding 
of the physical meaning of quantum and classical fermionic fields. 
However, we emphasize the practical utility of our 
method because the functional integration over fermionic degrees 
of freedom is not only very similar to that for bosons, but 
the numerical computation is even 
much faster than that for bosons. This is because, 
for each bosonic degree of freedom, the eigenvalues
$q$ and $p$ can take any real value, so lattice calculations require
Monte-Carlo simulations with a large number of different real values.
On the other hand, it appears that, 
for each fermionic degree of freedom, $q$ and $p$ can take only two
different values.     

First, consider one fermionic degree of freedom described by the 
operators $\hat{\psi}$ and $\hat{\psi}^{\dagger}$ that satisfy the 
anticommutation relations
\begin{equation}\label{algeb}
\{\hat{\psi},\hat{\psi}\} = 
\{\hat{\psi}^{\dagger},\hat{\psi}^{\dagger}\} =0, \;\;\;\;
\{\hat{\psi},\hat{\psi}^{\dagger}\} =1.
\end{equation}
This algebra can be represented on a 2-dimensional 
unitary space spanned 
by the orthonormal basis vectors $|0\rangle$ and $|1\rangle$ that 
satisfy $\hat{\psi}|0\rangle =0$,  
$\hat{\psi}^{\dagger}|1\rangle =0$, 
$\hat{\psi}^{\dagger}|0\rangle =|1\rangle$, and 
$\hat{\psi}|1\rangle=|0\rangle$.
The operators $\hat{\psi}$ and $\hat{\psi}^{\dagger}$ can be 
written as
\begin{equation}\label{psi}
\hat{\psi}=\hat{q}+i\hat{p}, \;\;\;\;
\hat{\psi}^{\dagger}=\hat{q}-i\hat{p},
\end{equation}
where
\begin{equation}\label{qp}
\hat{q}=(\hat{\psi}^{\dagger}+\hat{\psi})/2, \;\;\;\;
\hat{p}=i(\hat{\psi}^{\dagger}-\hat{\psi})/2.
\end{equation}
are hermitian operators. They satisfy the anticommutation relations 
\begin{equation}\label{algebqp}
\{\hat{q},\hat{q}\} =\{\hat{p},\hat{p}\}=1/2, \;\;\;\;
\{\hat{q},\hat{p}\}=0.
\end{equation}
In particular, we see that $\hat{q}^2=\hat{p}^2=1/4$, which implies 
that the only eigenvalues of $\hat{q}$ and $\hat{p}$ are 
$q_{\pm}=\pm 1/2$ and $p_{\pm}=\pm 1/2$, respectively. The corresponding 
eigenvectors are
\begin{equation}\label{eigenvectors}
|q_{\pm}\rangle =(|0\rangle \pm |1\rangle)/\sqrt{2}, \;\;\;\;
|p_{\pm}\rangle =(|0\rangle \pm i|1\rangle)/\sqrt{2},
\end{equation}
with the properties $\hat{q}|q\rangle =q|q\rangle$, 
$\hat{p}|p\rangle =p|p\rangle$, for $q$ ($p$) equal to 
$q_+$ or $q_-$ ($p_+$ or $p_-$). These eigenvectors satisfy
the orthonormality relations
\begin{equation}\label{orton}
\langle q| q'\rangle=\delta_{qq'}, \;\;\;\;
\langle p| p'\rangle=\delta_{pp'},
\end{equation}
and the completeness relations
\begin{equation}\label{complete}
\sum_q |q\rangle\langle q|=1, \;\;\;\;
\sum_p |p\rangle\langle p|=1.
\end{equation}
They also satisfy
\begin{equation}\label{projec}
\langle p|q\rangle =\frac{1}{2}\left(1-i4pq\right)=
\frac{1}{\sqrt{2}} e^{-i\pi pq}, 
\end{equation}    
which is similar to the analogous equation for bosons. (In 
order to obtain even closer similarity to bosons, one can 
remove the number $\pi$ in the last expression in (\ref{projec}) 
by an appropriate rescaling of the operators $\hat{q}$, 
$\hat{p}$, and their eigenvalues.)
 
Let us now apply the above equations to a calculation of the 
partition function $Z={\rm Tr}\,\exp(-\beta\hat{H})$, where 
$\hat{H}=\omega \hat{\psi}^{\dagger}\hat{\psi}$ is the Hamiltonian 
and $T=1/\beta$ is the temperature. Since 
$\hat{\psi}^{\dagger}\hat{\psi}|n\rangle =n|n\rangle$ for $n$ equal 
to 0 or 1, the simplest way to calculate $Z$ is as follows \cite{kap}:
$Z=\sum_n \langle n| \exp(-\beta\hat{H})|n\rangle =
\exp(-\beta\omega 0) + \exp(-\beta\omega 1) =
1+\exp(-\beta\omega)$. However, this simple method does not work 
for a more general case with many degrees of freedom and complicated 
interactions. Therefore, we present a different method, 
completely analogous to the functional method for bosons 
\cite{kap}, that can be generalized to more general cases.
We first write the partition function as 
\begin{equation}\label{Z}
Z=\sum_q \langle q|e^{-\beta\hat{H}}|q\rangle=
\sum_q \langle q| \left( \prod_{j=1}^{N} e^{-\Delta\tau\,\hat{H}} 
\right) |q\rangle ,
\end{equation}
where $\Delta\tau =\beta/N$. The next step is to insert the 
factor 1 in the forms of (\ref{complete}) $2N$ times and to put 
$N$ to infinity, so as to obtain
\begin{widetext}
\begin{eqnarray}\label{Z2}
Z & = &
 \lim_{N\rightarrow\infty}\sum_q
 \left(\prod_{j=1}^{N} \sum_{q(j)}\sum_{p(j)} \right) 
% \nonumber \\ & & \times
 \langle q |p(N) \rangle\langle p(N)|
 e^{-\Delta\tau\,\hat{H}}|q(N) \rangle\langle q(N) |p(N-1)\rangle
 \nonumber \\
 & & \times\langle p(N-1)|e^{-\Delta\tau\,\hat{H}}|q(N-1)\rangle
 \cdots 
% \nonumber \\ & & \times
 \langle q(2) |
 p(1) \rangle\langle p(1) |e^{-\Delta\tau\,\hat{H}}|q(1)\rangle\langle
 q(1) |q\rangle .
\end{eqnarray} 
Since 
the Hamiltonian is $\hat{H}=\omega \hat{\psi}^{\dagger}\hat{\psi}
=\omega(1/2-2i\hat{p}\hat{q})\equiv H(\hat{p},\hat{q})$ and  
since $\Delta\tau\rightarrow 0$, we can write
\begin{equation}\label{opnumb}
\langle p(j) |e^{-\Delta\tau\,H(\hat{p},\hat{q})}|q(j)\rangle
=
\langle p(j) |1-\Delta\tau\,H(\hat{p},\hat{q})|q(j)\rangle
%\nonumber \\
=
\langle p(j) |q(j)\rangle \, e^{-\Delta\tau\,H(p(j),q(j))}.
\end{equation}
\end{widetext} 
Therefore, using (\ref{opnumb}), (\ref{projec}),
and (\ref{orton}), (\ref{Z2}) can be written as
\begin{equation}\label{Z3}
Z  = 
 \lim_{N\rightarrow\infty}\sum_q
 \left(\prod_{j=1}^{N} \frac{1}{2}\sum_{q(j)}\sum_{p(j)} \right)
 \delta_{q(1),q} \, e^{-S},
\end{equation}
where
\begin{equation}\label{S}
S=\Delta\tau \sum_{j=1}^{N} \left[ -i\pi\frac{q(j+1)-q(j)}{\Delta\tau}
p(j)+H(p(j),q(j)) \right],
\end{equation}
and $q(N+1)\equiv q$. The summation over $q$ is easily performed, so 
the final expression is
\begin{equation}\label{Z4}
Z  =
 \lim_{N\rightarrow\infty}
 \left(\prod_{j=1}^{N} \frac{1}{2}\sum_{q(j)}\sum_{p(j)} \right)
 e^{-S},
\end{equation}  
where the periodic boundary condition $q(N+1)=q(1)$ is understood.
Similarity to the bosonic case \cite{kap} is obvious. 
As we have already discussed, the factor $\pi$ in (\ref{S}) can be 
eliminated by an appropriate rescaling. The 
only important difference is the fact that for bosons one has 
to make the replacement $(1/2)\sum_{q(j)}\sum_{p(j)}\rightarrow
(1/2\pi)\int dq(j) \int dp(j)$. Since for fermions
each $q(j)$ and $p(j)$ 
can take only two different values ($\pm1/2$), we see that 
the numerical computation (with a large but finite $N$)
for fermions is much faster than that 
for bosons. Note also that one can formally write (\ref{Z4}) 
with the continuum notation as 
$Z=\int\!\int{\cal D}q(\tau){\cal D}p(\tau)\,  e^{-S}$, where
$\tau=\beta j/N$, 
\begin{equation}
\int\!\int{\cal D}q(\tau){\cal D}p(\tau)\equiv
 \prod_{\tau=0}^{\beta} \frac{1}{2}\sum_{q(\tau)}\sum_{p(\tau)}, 
\end{equation} 
\begin{equation}\label{S2}
S=\int_0^{\beta} d\tau \left[ -i\pi\frac{dq(\tau)}{d\tau}
p(\tau)+H(p(\tau),q(\tau)) \right],
\end{equation}  
but here the derivative $dq(\tau)/d\tau\equiv (q(j+1)-q(j))/\Delta\tau$ 
is not well defined because it is always equal to 
$0$ or $\pm\infty$ as 
$\Delta\tau\rightarrow 0$. However, this is not a problem when 
$S$ is calculated as in (\ref{S}).

Let us now generalize the analysis to an arbitrary number of  
fermionic degrees of freedom. We label all degrees of freedom by 
$a=1,\ldots,L$, where $a$ comprises various types of degrees of 
freedom, such as different space points, spinor indices, and color 
indices. The anticommutation relations are
\begin{equation}\label{algebL}
\{\hat{\psi}_a,\hat{\psi}_b\} =
\{\hat{\psi}^{\dagger}_a,\hat{\psi}^{\dagger}_b\} =0, \;\;\;\;
\{\hat{\psi}_a,\hat{\psi}^{\dagger}_b\} =\delta_{ab}.
\end{equation}
We can represent this algebra in terms of direct products of 
representations for each pair $\hat{\psi}_a$, 
$\hat{\psi}^{\dagger}_a$. The orthonormal basis for the $2^L$-dimensional 
representation space can be chosen to be the set of $2^L$ states 
of the form
\begin{equation}
|n\rangle\equiv \bigotimes_{a} |n_a\rangle=
|n_1\rangle\otimes \cdots \otimes|n_L\rangle,
\end{equation}
where each $n_a$ is equal to 0 or 1. We take the action of the 
operators on these states to be defined by
\begin{equation}\label{order}
\hat{A}_1\cdots\hat{A}_L \,|n_1\rangle\otimes \cdots \otimes|n_L\rangle
=\hat{A}_1|n_1\rangle\otimes \cdots \otimes
\hat{A}_L|n_L\rangle,
\end{equation}
where $\hat{A}_a=A_a(\hat{\psi}_a,\hat{\psi}^{\dagger}_a)$ with 
$A_a$ being an arbitrary function. To determine the action of 
a differently ordered product of operators, one has to use the 
anticommutation relations (\ref{algebL}) to write the action in 
terms of operators ordered as in (\ref{order}). For example, 
$\hat{\psi}_2\hat{\psi}_1\,|n_1\rangle\otimes|n_2\rangle
=-\hat{\psi}_1\hat{\psi}_2\,|n_1\rangle\otimes|n_2\rangle
=-\hat{\psi}_1|n_1\rangle\otimes\hat{\psi}_2|n_2\rangle$. 
The hermitian conjugation of (\ref{order}) determines the 
action of the operators on the left:
\begin{equation}\label{order2}
\langle n_L|\otimes \cdots \otimes\langle n_1|\, 
\hat{A}^{\dagger}_L\cdots\hat{A}^{\dagger}_1=
\langle n_L|\hat{A}^{\dagger}_L\otimes \cdots \otimes
\langle n_1|\hat{A}^{\dagger}_1.
\end{equation}

To construct the functional formalism, we write 
$\hat{\psi}_a=\hat{q}_a+i\hat{p}_a$, so that $\hat{q}_a$ and 
$\hat{p}_a$ are hermitian operators that satisfy
\begin{equation}\label{algebqpL}
\{\hat{q}_a,\hat{q}_b\} =\{\hat{p}_a,\hat{p}_b\}=(1/2)\delta_{ab}, 
\;\;\;\; \{\hat{q}_a,\hat{p}_b\}=0.
\end{equation}
We introduce the states 
$|q\rangle= \bigotimes_{a} |q_a\rangle$ and 
$|p\rangle= \bigotimes_{a} |p_a\rangle$, which have the 
properties $\hat{q}_a|q\rangle=q_a|q\rangle$ and
$\hat{p}_a|p\rangle=p_a|p\rangle$.
These states satisfy the orthonormality and completeness relations 
analogous to (\ref{orton}) and (\ref{complete}), respectively, 
while the generalization of (\ref{projec}) is 
\begin{equation}\label{projecL}
\langle p|q\rangle =
\frac{1}{\sqrt{2^L}} e^{-i\pi pq},
\end{equation}
where $pq\equiv\sum_a p_a q_a$.

We want to calculate the partition function for a Hamiltonian of the 
form
\begin{equation}\label{H}
\hat{H}=\hat{\psi}^{\dagger}H^{(1)}\hat{\psi}+
(\hat{\psi}H^{(2)}\hat{\psi}+{\rm h.c.}),
\end{equation}
where $\hat{\psi}^{\dagger}H^{(1)}\hat{\psi}\equiv
\sum_{a,b}\hat{\psi}^{\dagger}_a
H^{(1)}_{ab}\hat{\psi}_b$, and similarly for the second term . We can write
\begin{equation}\label{H1}
H^{(1)}_{ab}=A_{ab}+S_{ab}, \;\;\;\; S_{ab}=S'_{ab}+\tilde{S}_{ab}, 
\end{equation}
where $A_{ab}=(H^{(1)}_{ab}-H^{(1)}_{ba})/2=-A_{ba}$ is the antisymmetric part, 
$S_{ab}=(H^{(1)}_{ab}+H^{(1)}_{ba})/2=S_{ba}$ is the symmetric part, 
$\tilde{S}_{ab}=\tilde{S}_a \delta_{ab}$, and $S'_{aa}=0$.
Using the anticommutation relations (\ref{algebqpL}), we can write 
\begin{eqnarray}\label{H11}
\sum_{a,b}\hat{\psi}^{\dagger}_a H^{(1)}_{ab}\hat{\psi}_b &=& 
(1/2)\sum_a \tilde{S}_a -2i\sum_a \hat{p}_a\tilde{S}_a\hat{q}_a
\nonumber \\
 & & +\sum_{a,b}\hat{q}_aA_{ab}\hat{q}_b 
+\sum_{a,b}\hat{p}_aA_{ab}\hat{p}_b.
\end{eqnarray}
We want the $\hat{q}_a$ operators to be ordered as in (\ref{order}), 
so we write
\begin{eqnarray}
\sum_{a,b}\hat{q}_aA_{ab}\hat{q}_b &=&
\sum_{a<b}\hat{q}_aA_{ab}\hat{q}_b + \sum_{a>b}\hat{q}_aA_{ab}\hat{q}_b
\nonumber \\
&=& \sum_{a<b}\hat{q}_aA_{ab}\hat{q}_b + \sum_{a<b}\hat{q}_bA_{ba}\hat{q}_a
\nonumber \\
&=& 2\sum_{a<b}\hat{q}_aA_{ab}\hat{q}_b 
\nonumber \\
&=& 2\sum_{a,b}\hat{q}_a A^<_{ab}\hat{q}_b,
\end{eqnarray}
where
\begin{equation}
A^<_{ab}=\left\{
 \begin{array}{l}
  A_{ab}, \;\; a<b, \\
  0,  \;\; a>b,
 \end{array}\right.
\;\;\;\;
A^>_{ab}=\left\{
 \begin{array}{l}
  0, \;\; a<b, \\
  A_{ab},  \;\; a>b.
 \end{array}\right.
\end{equation}
Similarly, we want the $\hat{p}_a$ operators to be ordered as in
(\ref{order2}), so we write
\begin{equation}
\sum_{a,b}\hat{p}_aA_{ab}\hat{p}_b=
2\sum_{a,b}\hat{p}_a A^>_{ab}\hat{p}_b .
\end{equation} 
We also apply a similar procedure for the second term in (\ref{H}). 
Without losing on generality, 
the matrix $H^{(2)}$ is taken to be antisymmetric. Defining the 
matrices $A^{(+)}$ and $A^{(-)}$ as 
$A^{(\pm)}_{ab}=H^{(2)}_{ab}\pm H^{(2)*}_{ab}$,   
we can write the Hamiltonian (\ref{H}) in the final form as 
\begin{eqnarray}\label{Hf}
H(\hat{p},\hat{q}) &=& \frac{1}{2}{\rm
Tr}\tilde{S}-2i\hat{p}\tilde{S}\hat{q}+
2\hat{q}A^<\hat{q}+2\hat{p}A^>\hat{p}
\nonumber \\
& & +2\hat{q}A^{(-)<}\hat{q}-2\hat{p}A^{(-)>}\hat{p}
+2i\hat{p}A^{(+)}\hat{q}.
\end{eqnarray}
This form of the Hamiltonian has the property
\begin{equation}
\langle p|H(\hat{p},\hat{q})|q\rangle =\langle p|q\rangle \, H(p,q).
\end{equation}
Therefore, we can repeat the steps that led to (\ref{Z4}) to obtain  
\begin{equation}\label{Z4L}
Z  =
 \lim_{N\rightarrow\infty}
 \left(\prod_{j=1}^{N} \frac{1}{2^L}\sum_{q(j)}\sum_{p(j)} \right)
 e^{-S},
\end{equation}
where $S$ has the form (\ref{S}) with the Hamiltonian $H(p,q)$ given
by (\ref{Hf}).  
In (\ref{Z4L}), $\sum_{q(j)}=\prod_a \sum_{q_a(j)}$, 
$\sum_{p(j)}=\prod_a \sum_{p_a(j)}$. 

Needless to say, the formalism developed above can also be applied to 
the calculation of the partition function at finite chemical potential. 
Moreover, essentially the same formalism can be applied to the calculation 
of correlation functions among fields at different times, in a way similar to 
that for bosons. Using the fact that $\hat{H}$ is the generator 
of infinitesimal time translations, one finds similar expressions 
with $\Delta\tau\rightarrow i\Delta t$. 

Concerning the application of our formalism to concrete physical models,
two additional remarks may be in order. First, one should be careful when the 
Hamiltonian is written in terms of Dirac fields $\hat{\psi}^{{\rm D}}$,
$\hat{\psi}^{{\rm D}\dagger}$. For example,  
if there are two degrees of freedom, one corresponding to a
particle and the other to an antiparticle, then the
``Dirac field" is
$\hat{\psi}^{{\rm D}}=(\hat{\psi}_1 + \hat{\psi}^{\dagger}_2)/\sqrt{2}$.
Second, although the
operator $\hat{N}=\sum_a \hat{\psi}^{\dagger}_a \hat{\psi}_a$ has the
property $\hat{N}|n\rangle =(\sum_a n_a)|n\rangle$, the
operator of the physical number of particles, in general,
has a more complicated form
$\hat{N}_{{\rm phys}}=\sum_{a,b} \hat{\psi}^{\dagger}_a
N_{ab}\hat{\psi}_b$. For the explicit
form of $N_{ab}$ in the continuum case, see \cite{nikol}.

In this Letter, we have presented a general formalism rather than 
apply it to concrete physical problems. However, it is straightforward 
to apply the developed formalism to any physical model, such as lattice QCD.
We strongly believe that this formalism will be widely applied to 
lattice QCD and lead to a number of important phenomenological results 
that previously could not be obtained because they
required too much computer time with previous algorithms for 
fermions.

Finally, let us shortly discuss the representation of quantum fermionic 
states in terms of wave functions $\Psi(q)$ written in the $q$-representation. 
Note that it is already 
known how to represent fermionic quantum states in terms of 
wave functions $\Psi(\psi)$, where $\psi$ is a Grassmann variable
\cite{flor}. However, $\Psi(\psi)$ is a Grassmann valued quantity, 
so $\Psi^*(\psi)\Psi(\psi)$ cannot be interpreted as a probability 
density. This problem does not exist for $\Psi(q)$. 
We first introduce the base functions
\begin{equation}\label{Psi}
\Psi_+(q)=\langle q|q_+\rangle =\Theta(q), \;\;\;\;
\Psi_-(q)=\langle q|q_-\rangle =\Theta(-q),
\end{equation}
where $\Theta(q)$ is the step function equal to 1 (0) for $q=1/2$ 
($q=-1/2$). From (\ref{qp}) and (\ref{eigenvectors}) we see that 
$\hat{p}|q_{\pm}\rangle =\mp(i/2)|q_{\mp}\rangle$. Therefore, 
we represent the operators $\hat{q}$ and $\hat{p}$ as 
\begin{equation}
\hat{q}=q, \;\;\;\; \hat{p}=-iq\hat{\epsilon}_q, 
\end{equation}
where $\hat{\epsilon}_q$ is the operator that acts on the space of 
functions as $\hat{\epsilon}_q f(q)=f(-q)$. We see that the operator 
$q\hat{\epsilon}_q$ plays the same role for fermions as the derivative 
operator $\partial_q$ does for bosons. The operators (\ref{psi}) 
are now represented as
\begin{equation}
\hat{\psi}=q(1+\hat{\epsilon}_q), \;\;\;\;
\hat{\psi}^{\dagger}=q(1-\hat{\epsilon}_q).
\end{equation}
It is easy to show that, for example, $\{\hat{p},\hat{p}\}f(q)=2q^2f(q)$, 
$\{\hat{\psi},\hat{\psi}^{\dagger}\}f(q)=4q^2f(q)$, which is 
in agreement with the anticommutation relations (\ref{algebqp}) and 
(\ref{algeb}), having in mind that the operators act on functions 
$f(q)$ defined for $q\in\{-1/2,1/2\}$. For time dependent wave functions, 
i.e., in the Schr\"odinger picture, the Schr\"odinger equation is 
\begin{equation}
H(-iq\hat{\epsilon}_q,q)\Psi(q,t)=i\partial_t \Psi(q,t).
\end{equation}
The expected value of an arbitrary operator $\hat{O}=O(\hat{q},\hat{p})$ 
is given by
\begin{equation}\label{O}
\langle \hat{O}(t)\rangle =\sum_q \Psi^*(q,t)O(q,-iq\hat{\epsilon}_q)
\Psi(q,t).
\end{equation}  
The generalization of Eqs.~(\ref{Psi})-(\ref{O}) 
to the case with many fermionic degrees of freedom is obvious.

The author is grateful to N.~Bili\'c
and H.~\v Stefan\v ci\'c for critical reading of the manuscript. 
This work was supported by the Ministry of Science and Technology of the
Republic of Croatia under Contract No.~0098002.

\end{document}